\documentclass[prd,twocolumn,floats,superscriptaddress,eqsecnum,floatfix]{revtex4}

\usepackage{amsmath}
\usepackage{graphics, graphicx}
\usepackage{epsfig}
\usepackage{ulem}
\usepackage{color}

\def\cO{{\cal O}}

\def\1p{{(1p)}}
\def\be{\begin{equation}}
\def\ee{\end{equation}}
\def\beq{\begin{eqnarray}}
\def\eeq{\end{eqnarray}}
\def\qf{}

\def\p0{\phi_0}
\def\z0{\zeta_0}

\def\aoi{D^{\ge 1}}

\def\Co{C_\ell^{\rm obs}}
\def\CK{C_\ell^K}
\def\Ct{C_\ell^{\rm tbo}}
\def\c{{\rm ei}}
\def\cob{}
\def\cor{}
\def\corr{}
\def\corb{}
\def\corrr{}
\def\corc{}
\def\crc{}
\def\qf{}
\def\qqf{}
\def\qr{}

\newcommand{\ttle}[1]{}

\begin{document}

\title{Local Observation in Eternal inflation}

\author{James  Hartle}
\affiliation{Department of Physics, University of California, Santa Barbara,  93106, USA}
\author{S.W. Hawking}
\affiliation{DAMTP, CMS, Wilberforce Road, CB3 0WA Cambridge, UK}
\author{Thomas Hertog}
\affiliation{APC, UMR 7164 (CNRS, Universit\'e Paris 7), 10 rue A.Domon et L.Duquet, 75205 Paris, France\\ {\it and}\\
International Solvay Institutes, Boulevard du Triomphe, ULB -- C.P. 231, 1050 Brussels, Belgium}

\bibliographystyle{unsrt}

\begin{abstract}

We consider landscape models that admit several regions where the conditions for eternal inflation hold.
It is shown that one can use the no-boundary wave function to calculate small departures from homogeneity within our past light cone despite the possibility of much larger fluctuations on super horizon scales.  The dominant contribution comes from the history exiting eternal inflation at the lowest value of the potential. In a class of landscape models this predicts a tensor to scalar ratio of about 10\%. In this way the no-boundary wave function defines a measure for the prediction of local cosmological observations. 

\end{abstract}


\maketitle

The string landscape is thought to contain a vast number of vacua, including some that {\cob have four large dimensions, our small positive value of {\corr the} cosmological constant, and the Standard Model.} But the landscape does not explain which {\cob vacuum in this class} we are in.  For that one has to turn to cosmology and to a theory of the quantum state of the universe.

A quantum state specifies amplitudes for different geometry and field configurations on a spacelike surface.
We have shown that the no-boundary wave function (NBWF) \cite{HH83} in the saddle point approximation predicts a large amplitude for configurations that behave classically when the universe is large and have an early period of inflation \cite{HHH08a}. The NBWF thus acts as a vacuum selection principle {\cob in the class described above, selecting} regions in field space where the {\cob landscape} potential admits one or more directions of inflation. The landscape then essentially becomes an ensemble of different models of inflation, weighted by {\corc NBWF probabilities}. {\cob For the rest of this paper we assume this ensemble.}

{\qf We are interested in the probabilities predicted by the NBWF for observables in our Hubble volume such as those of the cosmic microwave background (CMB). As observers we are physical systems within the universe that are described by {\qqf local} data $D$ that  include a specification of our observational situation. The data $D$ may occur in any Hubble volume with a small probability $p_E(D)$. Probabilities for our observations are NBWF probabilities conditioned on the requirement that at least that part of our data specifying the observational situation exist somewhere in the universe.  We call these conditional probabilities top-down (TD) probabilities to distinguish them from bottom-up (BU) probabilities conditioned only on the NBWF  \cite{HHe06,HH09}. }

We will find that significant contributions to TD probabilities come only from landscape regions that admit a regime of eternal inflation {\corrr where $V > \epsilon$}. {\cor Further it turns out that} the dominant contribution to TD probabilities comes from the region(s) where the threshold for eternal inflation lies at the lowest value of the potential, independently of its shape above this value.

In the usual approach to eternal inflation it is argued the universe develops large inhomogeneities that lead to a mosaic structure on super-horizon scales, consisting of (possibly infinitely many)  nearly homogeneous patches separated by inflating regions \cite{EIrefs}. The probability distributions for local observables can be different in different homogeneous patches, each of which itself can become arbitrarily large. This has led to a challenge, known as the measure problem, for the prediction of local observations in one Hubble volume. To resolve this a cutoff is imposed on the spacetime in order to regulate infinities. The expected number of Hubble volumes of different kinds can then be calculated and used to define the probabilities for the observations of a typical observer. 

{\corc A  very different approach to eternal inflation is {\crc based on} the measure defined by the universe's quantum state.} {\crc This paper builds on a series  (e.g. \cite{HHH08a,HHH10a}) in which} we have investigated the {\crc implications of the }no-boundary quantum state measure.
The semiclassical NBWF  does not predict a single classical spacetime. Rather it predicts {\cob BU probabilities for an ensemble of alternative spacetimes}. In models of eternal inflation the  {\corc TD probabilities} for large long wavelength perturbations are high \cite{HHH10a}. 
However, in contrast to the usual approach,   we find that the very large scale structure of the eternally inflating histories in the ensemble is irrelevant for the probabilities of observables in our Hubble volume. The latter depend only on alternatives in our past light cone.  To calculate the {\corc probabilities}  of different configurations inside one Hubble volume one sums (coarse grains) over everything outside the past light cone.  This results in well-defined probabilities for observations without the need for further ad hoc regularization. 


{\bf The No-Boundary Measure} 

A quantum state of the universe is specified by a wave function $\Psi$ on the superspace of  geometries ($h_{ij}(x)$)  and matter field configurations ($\chi(x)$) on a closed spacelike {\cob three}-surface $\Sigma$. Schematically we write $\Psi=\Psi[h,\chi]$. We assume the no-boundary wave function as a model of this state \cite{HH83}. The NBWF is given by a sum over histories of geometry $g$ and fields $\phi$ on a four-manifold with one boundary $\Sigma$. The contributing histories match the values $(h,\chi)$ on $\Sigma$ and are otherwise regular. They are weighted by $\exp(-I/\hbar)$ where $I[g,\phi]$ is the Euclidean action.

In some regions of superspace the path integral can be approximated by the method of steepest descents. Then the NBWF will be approximately given by a sum of terms of the form 
\begin{equation}
\Psi[h,\chi] \approx  \exp\{(-I_R[h,\chi] +i S[h,\chi])/\hbar\} ,
\label{semiclass}
\end{equation}
one term for each complex extremum.  Here $I_R[h,\chi]$ and $-S[h,\chi]$ are the real and imaginary parts of the Euclidean action, evaluated at the extremum. 

When the surfaces $\Sigma$ are three spheres of radius $a$ with $a^2 V(\phi) <1$, where $V$ is the potential of the scalar matter fields, there is an approximately real Euclidean solution of the field equations, and $S\approx 0$. 
For large radii $a$, however, there are only complex solutions, and the wave function oscillates rapidly. When $S$ varies rapidly compared to $I_R$ (as measured by quantitative classicality conditions \cite{HHH08a}) the NBWF predicts that  the geometry and fields behave classically. The {\corc NBWF} can then be viewed as predicting a family of classical Lorentzian histories that are the integral curves of $S$ and have probabilities to leading order in $\hbar$ that are proportional to  $\exp[-2 I_R(h,\chi)]/\hbar]$, which is constant along the integral curve.

In \cite{HHH08a,HHH10a} we evaluated the semiclassical NBWF  for a model consisting of a single scalar field moving in a quadratic potential. {\corc We found}  that for large radii $a$ the NBWF predicts a family of alternative Lorentzian Friedman-Lema\^itre-Robertson-Walker universes with Gaussian perturbations. The alternative histories can be labeled by the {\cor absolute values} of the perturbations $\zeta_0$ and {\cor of} the background scalar field  $\phi_0$
at the `South Pole'  (SP) of the corresponding saddle point. Classicality requires $\phi_0 \gtrsim 1$ (in Planck units). The relative BU probabilities of the alternative configurations follow approximately from
\be
I_R(\phi_0) \approx - \pi/4 V(\phi_0)
\label{nbweight}
\ee
together with the Gaussian probabilities for fluctuations. 

The NBWF has the striking property that all saddle point histories undergo some amount of matter driven slow roll inflation, with a number of $e$-folds  $N(\p0) \approx 3 \phi_0^2/2$. The NBWF therefore {\it selects} inflationary classical histories. These exhibit the usual Gaussian spectrum of fluctuation modes $\zeta_q$ with expected amplitude $\zeta_q^2  \approx (\pi^2/4)\left({H^2}/{\epsilon} \right)_{\rm exit}$, where $\epsilon \equiv \dot \phi^2 /H^2$ and the amplitude is evaluated when the perturbations exit the horizon. {\corrr Hence saddle points starting below the threshold of eternal inflation are nearly homogeneous. By contrast, in regions of the landscape where the condition for eternal inflation $V > \epsilon$ is satisfied -- which means the scalar fields are effectively in a deSitter background -- the {\corc probabilities} are high for significant perturbations on large scales.}

Assuming this holds generally,  only configurations that emerge from regions of the landscape {\crc that admit inflationary solutions} will have significant {\corc probability}. One expects different inflationary patches of a landscape potential are separated by large potential barriers. Hence the NBWF acts as a vacuum selection principle in the landscape which becomes an ensemble of models of inflation, weighted by their no-boundary {\corc probabilities}.


We now seek to calculate the relative contributions of the different inflationary regions in field space to the TD probabilities for local observations in our Hubble volume. 

{\bf Probabilities for Observations} 

{\it Top-Down Weighting:}
Our observations are confined to one Hubble volume and our data $D$ occur with only a very small probability $p_E (D)$ in any Hubble volume on a constant density surface $\Sigma_s(D,\p0,\zeta_0)$. {\corc TD} probabilities for local observables ${\cal O}$ {\crc takes this observational situation in account by weighting the BU} probabilities by the probability that $D$ exists somewhere on the surface.
We showed in \cite{HH09} that this  weighting is {\crc given by} a multiplicative factor
\begin{equation}
\label{tdweight}
 1-[1-p_E(D)]^{N_h} \leq 1. 
 \end{equation}
 Here, $N_h$ is the total number of Hubble volumes in $\Sigma_s$. When $N_h$ is sufficiently small so that our data is rare the TD factor is very small and reduces to weighting the BU probabilities by the volume of $\Sigma_s$ \cite{Page97,Hawking07}.  We have argued \cite{HH09}  this is the case in saddle point histories that start below the threshold of eternal inflation, which predict {\cor high amplitudes for} nearly homogeneous final configurations with {\crc \eqref{tdweight}} proportional to $e^{3N}$.

By contrast, saddle points in the regime of eternal inflation predict high amplitudes for configurations that have large long wavelength perturbations \cite{HHH10a}. The volume of $\Sigma_s$ of these perturbed configurations can be exceedingly large or even infinite \cite{CREM} so that the probability that $D$ exists somewhere \eqref{tdweight} is nearly one \footnote{{\qqf We do not consider Boltzmann brain instances of D, which have lower probability in the saddle point approximation}.}.

{\cor The TD weighting has {\cob a significant} effect on the BU distributions in models that admit a regime of eternal inflation. The NBWF BU probabilities favor histories starting at a low value of the potential {\cob followed by} only a few $e$-folds of slow roll inflation [cf. \eqref{nbweight}]. However, the {\cob TD weighting \eqref{tdweight}  suppresses the probabilities for such} histories, and instead favors saddle points starting in the regime of eternal inflation}. 
{\crc In models of eternal inflation, the low BU probability of histories starting above the threshold of eternal inflation is compensated by the large number of Hubble volumes in the resulting surfaces  $\Sigma_s$.}
{\cob  Once regions in field space for which the condition for eternal inflation holds have been selected}, the probabilities of local observations are given by their bottom-up values. It remains to estimate the latter.
\\{\qf {\it Cosmic No Hair:}   For observations we only need the NBWF probabilities for fluctuations in our Hubble volume located somewhere on the reheating surface $\Sigma_s$. That surface will be very large and inhomogeneous as a consequence of fluctuations that left the horizon during eternal inflation. The cosmic no hair theorems imply that predictions on the scales of a Hubble volume will be independent of the detailed large scale structure of $\Sigma_s$  provided that there are a sufficient number of efolds after the exit from eternal inflation. Rather, the predictions on Hubble volume scales are the same as in a homogeneous universe perturbed by small fluctuations.  
   
Usual derivations of cosmic no-hair results assume the Bunch-Davies vacuum for subhorizon modes at the start of inflation. That assumption is replaced here by the NBWF. Its predictions for fluctuations on Hubble volume scales were calculated for $(1/2)m^2\phi^2$ potentials in \cite{HHH10a}. Further for realistic values of $m$ the NBWF histories have the necessary large number of efolds ($\sim 1/m)$ after exit from eternal inflation.}
\\ {\it Coarse Graining:}
{\qf These results can be derived explicitly from the NBWF by coarse graining. }
A wave function $\Psi[h,\chi]$, and the histories constructed from it, specify probabilities for alternatives all across a spacelike surface, to its future, and to its past.
Probabilities for observables $\cO$ in our Hubble volume are obtained by summing the NBWF probabilities over alternatives off that surface and outside that Hubble volume. This is coarse graining. 

Causality implies that the probabilities of observations in our Hubble volume can depend only on alternatives in our past light cone. Coarse graining over alternatives to the future of $\Sigma_s$, including all quantum branching, is immediate.  {\qf Probabilities for alternatives on $\Sigma_s$ are then related directly to the wave function on $\Sigma_s$.} 

There remains the coarse graining over alternatives on $\Sigma_s$ in Hubble volumes outside our own. This can be discussed explicitly in the saddle point approximation \eqref{semiclass}. {\qf The cosmic no-hair results imply} 
that a given saddle point yields the same predictions for local observables for all Hubble volumes 
on $\Sigma_s$. {\cor Further, all saddle points starting at sufficiently high potential in a particular inflationary direction in field space yield identical predictions. This is the case, for instance, in single-field models of eternal inflation with a sufficient number of $e$-folds after the exit from eternal inflation.} 
We call regions in the landscape where saddle points produce {\corr Hubble volumes with the same distributions for $\cal O$} eternally inflating {\it channels}.  Probabilities for {\cob observables $\cal O$} depend only on the channel, and we can coarse grain over different eternally inflating histories in any one channel. Labeling the different channels by $K$ we arrive at
\be
\label{FF}
p({\cal O}|\aoi) \approx \sum_{K} p({\cal O}|K) p(K).
\ee
Here, $p(K)$ is the NBWF probability  of channel $K$, and $p({\cal O}|K)$ is the probability for observables $\cO$ given that channel. {\qf In this approximation the TD probabilites $p({\cal O}|\aoi)$ are independent of the details of the data $D$. } This is a finite and manageable prescription for probabilities for observation in our Hubble volume. {\cor The two probabilities involved in \eqref{FF} can be estimated as follows.}

{\cor Local observables $\cO$ such as those associated with the CMB refer only to short wavelength fluctuations that can be observed in our Hubble volume. This means} long-wavelength fluctuations should be coarse grained over to compute $p(\cO,K)$. {\corrr To leading order in $\hbar$}, if one coarse grains over all possible values of the long-wavelength fluctuations this sum-over-histories yields one. The probabilities $p(\cO,K)$ can then be estimated by using saddle points in channel $K$ that are nearly homogeneous everywhere and retain only the small, 
short-wavelength, observable, fluctuations. {\cor Those saddle points can be labeled by {\corc $\phi_{K0} \geq \phi^\c_K$, }where $\phi^\c_K$ is the threshold value that marks the onset of the regime of eternal inflation in channel $K$, and by the values of the short-wavelength perturbations {\corc $\zeta_{K0}$} at the SP. Since the BU probabilities decrease rapidly with $\p0$ (cf. \eqref{nbweight}) we can approximate $p(K)$ by $\exp(-\pi/(4V_K^\c))$ where $V_K^\c \equiv V(\phi^\c_K)$ is the value of the local potential in channel $K$ at the lowest exit from eternal inflation. {\cor The conditional probabilities $p(\cO |K)$ can be calculated using standard perturbation theory techniques (see e.g. \cite{HHH10a}).} 

{\cor Hence, with these approximations we predict that the contributions from different channels in the landscape to the TD probabilities \eqref{FF} of observables $\cO$ in our Hubble volume are approximately given by the homogeneous saddle {\cob points with  the lowest exits} from eternal inflation in each channel. 
{\corr If the landscape has one particular channel where the threshold of eternal inflation is at significantly lower potential than in all others, then this channel provides the dominant contribution to the sum in \eqref{FF}.}}

{\bf Predictions for Observations}

{\sl Models of Inflation:}
{\corrr So far we have concentrated on explaining how predictions for observations can be derived from a quantum state of the universe in a landscape that allows for {\crc different} regions of eternal inflation. However, one cannot expect this general discussion to yield realistic predictions without further qualification. In particular we have not discussed possible structure on the landscape, nor optimized the class of data $D$ assumed for TD probabilities. 

Therefore to illustrate the framework above we now consider a model} landscape where the NBWF selects a discrete set of $K$ minima that are separated from each other by steep potential barriers. {\crc We further} assume that each {\cor minimum {\cob in the class under discussion} has} a single inflationary direction $\phi_K$ in field space, where the local potential $V(\phi_K) = \mu_K\phi_K^{n_K}$ with $n_K\geq 2$, {\crc and} that otherwise the minima are similar. For simplicity we assume the value of $\mu_K$ agrees with the amplitude measured by Cosmic Background Explorer (COBE). The threshold values $\phi_K^\c$ that mark the onset of the regime of eternal inflation around the different minima {\cor can be calculated from the condition that $V^3= V_{,\phi}^2$ at $ \phi_K^\c$.}
Substituting the NBWF {\corc probabilities} (cf. \eqref{nbweight}) {\cor with $\phi_0 = \phi_K^\c$} in \eqref{FF} yields for the TD probabilities
\be
p ({\cal O}|\aoi) \approx \sum_{K} p({\cal O}|\phi_K^\c) \exp\left({1}/{\mu_K}\right)^{\frac{2}{2+n_K}} \ .
\label{ptd}
\ee
Hence, {\qr with the assumed $\mu_K$,}  the dominant contribution comes from minima where the scalar field is moving in a quadratic potential, for which $p(\phi_K^\c) \approx \exp(1/m)$ where $m=\sqrt{\mu}$.  {\qr Then we predict that the observed CMB temperature anisotropies will be those of an inflationary model with a quadratic potential. Specifically, in this model landscape, we predict an essentially Gaussian spectrum of microwave fluctuations with a scalar spectral index $n_s \sim .97$ and a tensor to scalar ratio of about 10\% \cite{Weinberg}.}

{\sl Contributions from Different Saddle Points:} 
The NBWF predicts probabilities $p(\Co|K)$ for the standard  multipole coefficients of the observed CMB two point correlator in any class of backgrounds $K$ \cite{HHH10a}. The BU NBWF probabilities for fluctuations are Gaussian to lowest order in their amplitude. The resulting probabilities $p^\ell(\Co|K)$ for a given $\ell$ are therefore {\corb essentially} a $\chi^2$-distribution specified by a mean $\langle \Co \rangle = \CK $ and (cosmic) variance  $\sigma^K_\ell \equiv 2(\CK)^2/(2\ell+1)$ 
where the $\CK$ are the theoretical multipole coefficients that completely characterize Gaussian fluctuations.  

If the $\CK$ differ significantly we would expect all $\Co$'s to be within a few $\sigma$'s of one {\it or} the other predicted expected values $\CK$. 
If  some $\Co$'s have been measured, they can be used  to make predictions about $\Ct$'s to be {\corc observed} by computing the conditional probability $p(\Ct|\Co)$. Since the $\Co$'s are independent random variables these turn out to be given by
\begin{equation}
p(\Ct|\Co)= \sum_K p(\Ct|K)p(K|\Co)  
\label{CfrmC}
\end{equation}
where, using the Bayes relation, 
\begin{equation}
p(K|\Co)= \frac{p(\Co|K)p(K)}{\sum_Kp(\Co|K)p(K)} 
\label{KfrmC}
\end{equation}
and $p(K)$ is  the NBWF probability for {\cor channel} $K$.
If {\cor either the no-boundary {\corc probabilities} $p(K)$ or} the $\Co$'s are enough to make $p(K|\Co)$ peaked around one $K$ then \eqref{CfrmC} predicts that further observations will confirm that.

Finally we note that although the {\crc TD} probabilities {\crc \eqref{FF}} for {\cor linear} fluctuations are a sum of Gaussian distributions from different channels, no {\cor non-Gaussianity} is predicted for the standard {\cor measures of it} as discussed in \cite{HHH10a}.

{\bf Conclusion}

{\corc As a quantum mechanical system, the universe has a quantum state. A theory of that state such as the NBWF is a necessary part of any final theory. 
The probabilities following from the state are a measure for prediction in cosmology. Applied to predictions of our local observations the NBWF measure appears to be finite without the need for further ad hoc regularization. We briefly summarize the essential principles behind this.}

{\crc The state predicts probabilities for different configurations of geometry and field on a spacelike surface.}
Our observations of the universe are limited to one particular Hubble volume in a much larger universe.   {\corc Their probabilities are defined by summing (coarse-graining) over unobserved features}, for example the location of our past light cone in spacetime, or structure arising from quantum events far outside our past light cone. The saddle point approximation to the wave function incorporates some of this coarse graining. The resulting probabilities for observation are well-defined and depend only on alternatives in our past light cone.

Applying the {\corc no-boundary measure} to a model landscape we found the dominant contribution to top-down probabilities comes from the region(s) in field space where the threshold for eternal inflation holds at the lowest value of the potential. In {\qf the particular} model consisting of isolated minima with polynomial, monotonically increasing directions of inflation, this implies an essentially Gaussian spectrum of microwave fluctuations with a scalar spectral index $n_s \sim .97$ and a tensor to scalar ratio of about 10\%.

\noindent{\bf Acknowledgments:}  We thank B. Freivogel, A. Guth, D. Page, M. Salem, M. Srednicki and A. Vilenkin  for helpful discussions. {\crc JH and TH} thank the CTC at Cambridge for its hospitality. We thank Chris Pope, Sheridan Lorenz and the Mitchell Institute for several meetings at Cooks Branch.  The work of JH was supported in part by the US NSF grant PHY07-57035 and  by Joe Alibrandi. The work of TH was supported in part by the ANR (France) grant ANR-09-BLAN-0157.

\eject

\end{document}